# Fibre-integrated van der Waals quantum sensor with an optimal cavity interface

*Jong Sung Moon[1,*], Benjamin Whitefield[2,3,*], Lesley Spencer[2,3,*], Mehran Kianinia[2,3], Madeline Hennessey[2,3], Milos Toth[2,3], Woong Bae Jeon[1], Je-Hyung Kim[1,^] and Igor Aharonovich[2,3,^]*

[1]Department of Physics, Ulsan National Institute of Science and Technology (UNIST), Ulsan 44919, Republic of Korea

[2]School of Mathematical and Physical Sciences, University of Technology Sydney, Ultimo, New South Wales 2007, Australia

[3]ARC Centre of Excellence for Transformative Meta-Optical Systems (TMOS), University of Technology Sydney, Ultimo, New South Wales 2007, Australia

 *equal contribution

^Corresponding author: jehyungkim@unist.ac.kr ; igor.aharonovich@uts.edu.au

**Abstract**

*Integrating quantum materials with fibre optics adds advanced functionalities to a variety of applications, and introduces fibre-based quantum devices such as remote sensors capable of probing multiple physical parameters. However, achieving optimal integration between quantum materials and fibres is challenging, particularly due to difficulties in fabrication of quantum elements with suitable dimensions and an efficient photonic interface to a commercial optical fibre. Here we demonstrate a new modality for a fibre-integrated van der Waals quantum sensor. We design and fabricate a hole-based circular Bragg grating cavity from hexagonal boron nitride (hBN), engineer optically active spin defects within the cavity, and integrate the cavity with an optical fibre using a deterministic pattern transfer technique. The fibre-integrated hBN cavity enables efficient excitation and collection of optical signals from spin defects in hBN, thereby enabling all-fibre integrated quantum sensors. Moreover, we demonstrate remote sensing of a ferromagnetic material and of arbitrary magnetic fields. All in all, the hybrid fibre-based quantum sensing platform may pave the way to a new generation of robust, remote, multi-functional quantum sensors.*

Optical fibres underpin many mature technologies ranging from high-speed internet to sensing and imaging. Integrating optical fibres with active functional elements underpins future-generation remote sensing devices, advanced biophotonic and photonic technologies, and enables new modalities for light emitting diodes, lasers and non-linear optics[1-6]. For these reasons, fibre sensors are already used widely in numerous fields, ranging from inertial navigation[7] to medical diagnosis[8] . In the ever-evolving realm of quantum sensing, there is a growing demand for integrating active quantum resources into highly-matured fibre optic platforms. Whilst laser coupling into a fibre is straightforward, achieving efficient excitation and optical readout of an integrated sensor is not, as it requires both mode matching as well as submicron alignment between the functional component (e.g., a quantum emitter) and the fibre[9, 10].

To date, substantial attention has been given to quantum sensors based on the nitrogen-vacancy (NV) centre in diamond[11, 12]. However, the integration of bulk diamond or nanodiamonds with fibres is challenging for several reasons. First, the NV centre emission overlaps spectrally with fibre autofluorescence, which prohibits efficient sensor operation[13]. Second, integration of bulk crystals that possess high refractive index significantly limits the light extraction efficiency and applications that require nanoscale focusing or spatial resolution[14]. Alternatively, nanodiamonds attached to a fibre core result in inefficient excitation and significant losses at collection pathways, particularly at small numerical apertures (NAs)[10, 15]. A workaround is to use bulk optics for imaging or integrate nanostructures at the fibre tip, which requires cumbersome manipulation and often results in low coupling efficiencies[16-18]. These approaches limit the overall performance and utility of devices and consequently, efficient integration of solid-state quantum sensors and fibre optics has, to date, remained elusive.

The van der Waals material hexagonal boron nitride (hBN) has recently garnered significant attention due to its layered nature and unique optical properties[19, 20]. Importantly, it can host on-demand single photon emission, as well as various spin defects that are suitable for quantum sensing applications[21-27]. A major advantage of hBN is the ability to exfoliate thin flakes with controllable thickness, enabling the fabrication of nanophotonic cavities used to enhance light extraction and coupling[28-30]. Additionally, hBN flakes can be positioned easily and deterministically using well-established, mature flake transfer techniques[31]. As a result, hBN is an ideal material platform for hybrid-integrated quantum devices and, particularly appealing for integration with optical fibres.

Here, we introduce a new fibre-integrated hBN cavity, and use it to demonstrate proof-of-principle remote quantum sensing. Specifically, we fabricate a hole-based circular Bragg grating (CBG) cavity that generates a vertically narrow Gaussian beam profile from hBN. We subsequently engineer, on demand, spin defects (negatively charged boron vacancies, $V_B^-$) within the CBG, and transfer it onto the core of a commercial optical fibre. The engineered fibre cavities overcome the shortfalls of prior attempts by efficiently coupling the emission from $V_B^-$ centres into the fibre, and enable proof-of-principle remote sensing of magnetic fields. Our results constitute a new hybrid system for remote sensing, with unprecedented functionalities.

Figure 1(a) schematically illustrates the fibre-integrated quantum sensor consisting of a monolithic CBG hBN cavity integrated with a standard optical fibre. The CBG contains an ensemble of optically-active spin defects that can be excited and read out through the fibre. The inset depicts the $V_B^-$ spin defects in hBN which are used as sensors in this work. The CBG acts both as the cavity for light collection as well as the active medium for quantum sensing. The optical operation, including laser excitation and photoluminescence (PL) collection can all be done through fibre without any additional optical components.

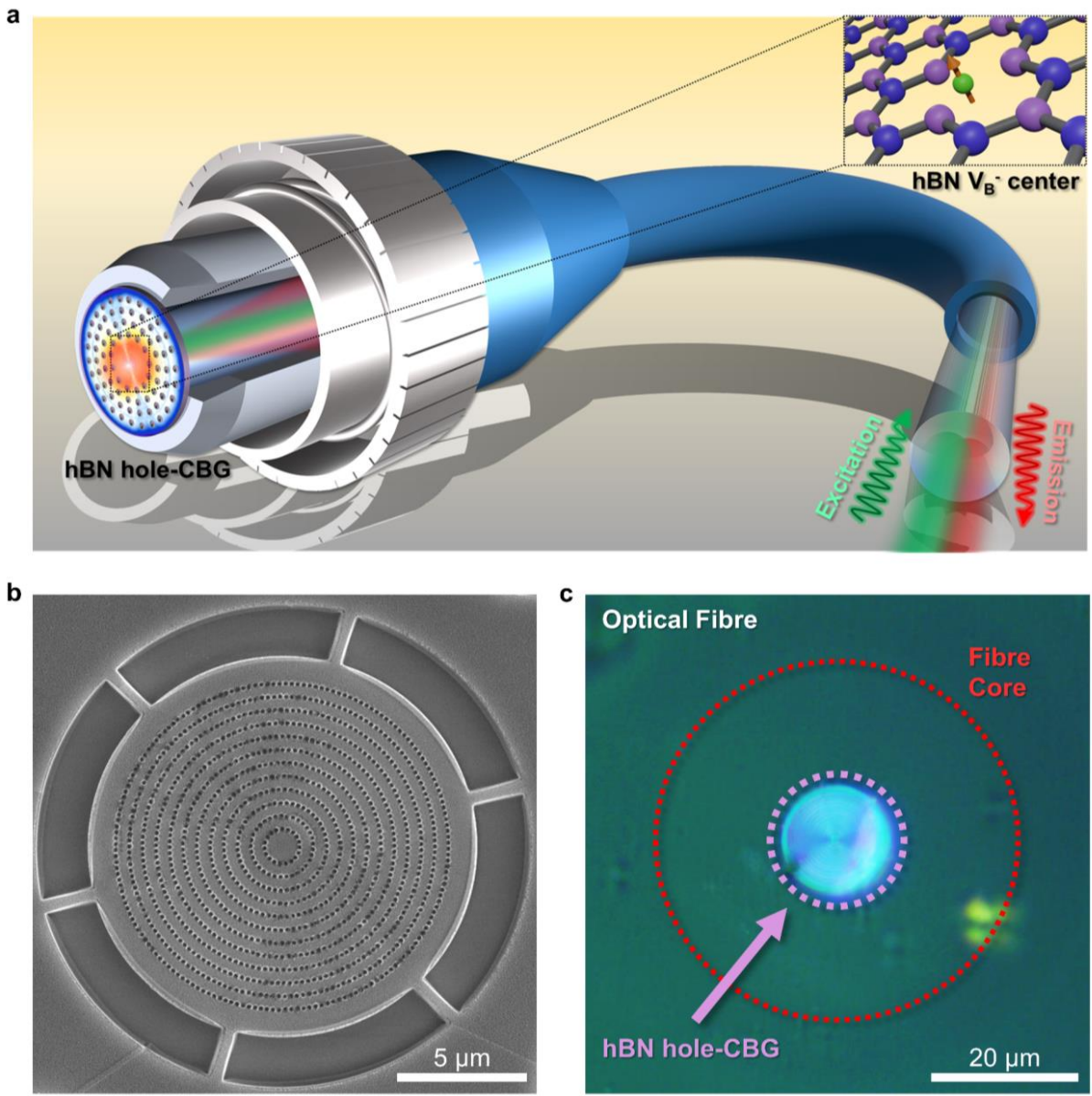

***Figure 1.*** **Fibre-integrated quantum sensor (a)** *Schematic illustration of the fibre-integrated quantum sensor. A hBN circular bragg grating (CBG) is placed at the end of the fibre core, enabling excitation and collection of hBN spin defects embedded in the CBG. The inset is a visualisation of the negatively charged boron vacancy, the spin-active defect used for sensing.* **(b)** *SEM image of the hBN CBG. The CBG is connected to the hBN flake with a sprue-like release structure that is broken when transferring the CBG onto the fibre.* **(c)** *Optical image of the hBN CBG cavity integrated onto the fibre core.*

A scanning electron microscopy (SEM) image of a fabricated CBG hBN cavity which is designed to enable ergonomic transfer onto a fibre is shown in Figure 1(b). The cavity is composed of a nanohole array that is periodic in both the radial and tangential directions. This design enhances the vertical directionality of the coupled emission compared to a conventional CBG composed of concentric rings[32]. The sprue-like release structure around the CBG provides a temporary mechanical connection to the parent hBN flake – it breaks when the CBG is detached from the flake and transferred to a fibre core. The transfer of the hBN CBG onto the fibre was done using a standard polydimethylsiloxane (PDMS) stamp two-dimensional (2D) material transfer technique. Figure 1(c) shows an optical image of the fibre-integrated CBG device post transfer. Details of the fibre integration process and the cavity fabrication method are described in Supplementary Information.

The CBG cavity used in this work was designed specifically to enable integration with a fibre, and to optimise fibre coupling. The design (termed "hole-CBG") replaces the air gaps between the rings of a conventional CBG with nanohole arrays between the rings[32]. This has two primary advantages. First, whilst all CBGs redirect emitted light in a vertical, out of plane direction, the hole-CBG achieves improved narrowing of the emission in the far field, yielding more efficient mode matching and fibre coupling. Second, the entire hole-CBG structure is interconnected structurally, which is critical for robust, reliable transfer of such cavities without geometric distortion and damage that would occur in conventional CBGs comprised of concentric rings separated by continuous air gaps. During CBG fabrication, the supporting sprue-like structures keep the cavity connected to the parent hBN flake until they are broken during the lift-out step of the transfer process. The cavity engineering process is assisted by the layered nature of hBN. First, the ability to exfoliate hBN enables ergonomic fabrication of the thin membranes used to etch CBGs. Second, it simplifies the transfer process, and enables transfer of CBGs onto fibre cores using a low-cost method that is rapid, reliable and can, in the future, be scaled and automated.

To optimise the cavity design, we performed a finite-difference time-domain (FDTD) simulation (See methods). An optimised cavity geometry has the following dimensions: a central disk radius ($R$) of ~ 645 nm, a radial interval between nanoholes arrays ($r$) of ~ 498 nm, a tangential interval between nanoholes ($l$) of ~ 224 nm , and a nanohole radius ($a$) of ~ 75 nm. Figure 2(a) shows the simulated far-field emission profile of the CBG cavity compared to that of a pristine hBN flake. The hBN CBG exhibits a highly-focused beam profile within a propagation angle of 10 degrees, whereas the pristine flake has a widespread emission angle. This simulation confirms that the designed cavities can be efficiently coupled to low-NA optical systems such as optical fibres. To compare the CBG and the pristine hBN flakes quantitatively we calculate, in Figure 2(b), the collection efficiency as a function of the numerical aperture. The approximate numerical apertures of single mode fibres, multimode fibres and high-NA optical systems are shaded yellow, green and blue, respectively. In the case of the hBN CBG, 50% collection efficiency can be achieved even at an NA of 0.1, corresponding to a single-mode fibre. A multimode fibre would result in over 80% collection efficiency employing the same CBG cavity. In contrast, emission from pristine hBN is barely coupled to a single mode fibre, and only 20% of the emission is coupled to a multi-mode fibre. The simulations illustrate the high efficiency of the optimised cavity design, and illustrate the ability to interface optimally with optical fibres.

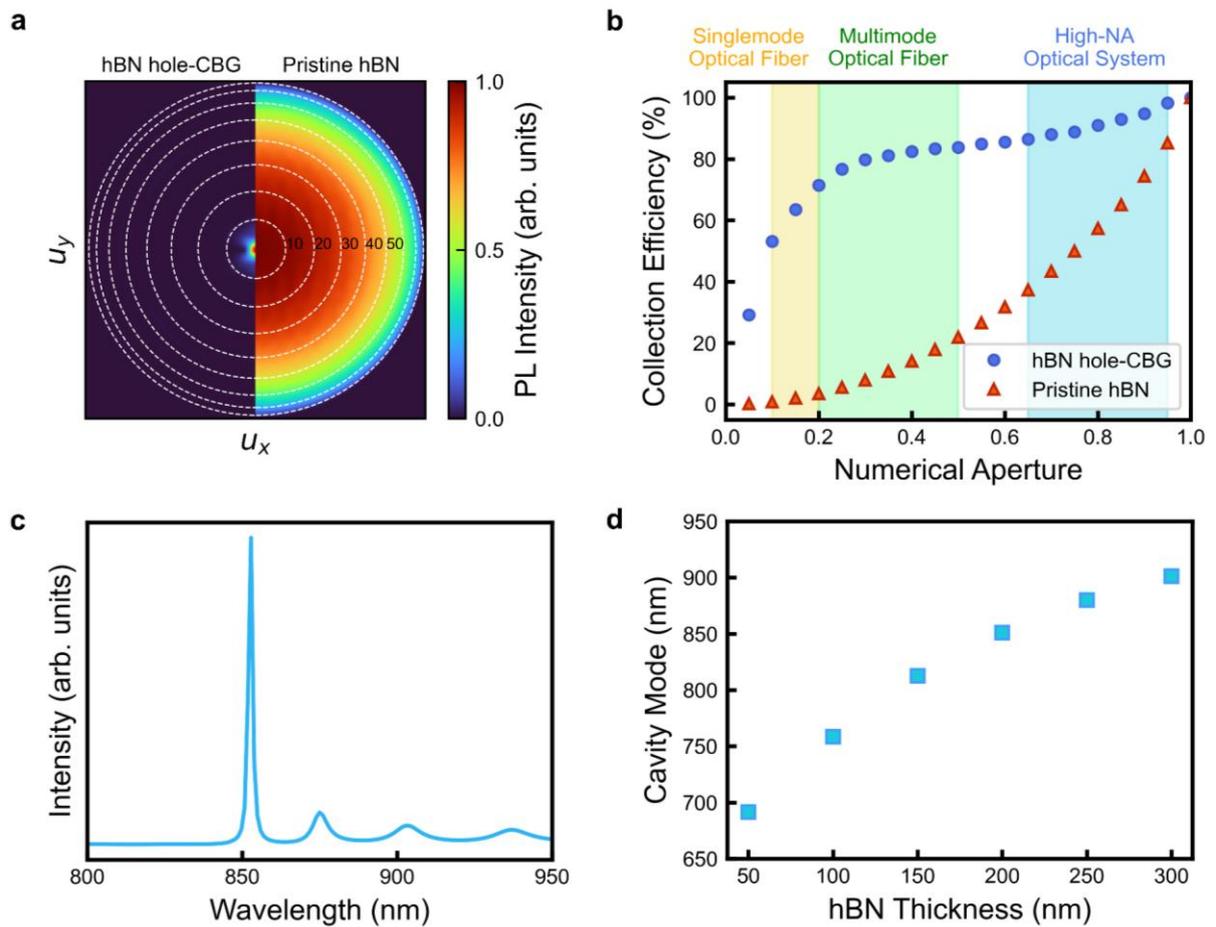

***Figure 2. FDTD design of hBN hole-CBG cavities. (a)** Far-field emission profile map of an optimised hBN hole-CBG (left) and a pristine hBN flake (right) **(b)** Comparison of the collection efficiency of the hBN CBG cavity (blue circles) and a pristine flake of hBN (red triangles) as a function of numerical aperture. The shaded regions (yellow, green, blue) denote the corresponding numerical aperture of a single mode fibre, multimode fibre and a high-NA optical system. **(c)** Simulated cavity mode spectrum in the emission range of the $V_B^-$ defect. **(d)** Central wavelength of the cavity mode as a function of the thickness of the original hBN flake.*

Figure 2(c) is a simulated mode spectrum of the optimised cavity design. It exhibits an intense, sharp resonance at 850 nm, to maximise the overlap with the $V_B^-$ spin defects which have a broad emission at ~ 750 - 900 nm[23]. The cavity mode indicates strong Purcell enhancement ($F_P$ ~ 50) which enhances the radiative decay rate of the emitters, and should increase the fluorescence intensity of the defects. We note that the choice of the $V_B^-$ spin defect offers a practical advantage in that the emission does not overlap spectrally with the fibre autofluorescence which dominates the red spectral range. Finally, Figure 2(d) shows tunability of the cavity mode as a function of the hBN thickness. As anticipated, the cavity mode wavelength red-shifts with increasing thickness of hBN.

Once the parameter space for the hBN cavities was established, the devices were fabricated by patterning the optimised cavity design with e-beam lithography followed by dry etching, and transferred onto optical fibres using a PDMS microstamp. We used a multimode fibre with a core size of 50 μm and NA=0.22. (See Supplementary Information for technical details). First, we characterised the performance of the CBG mounted onto the fibre core Figure 3(a, b) show the FDTD simulation of the near-field power ($|E|^2$) map and the corresponding measured PL map of the hBN CBG cavity, respectively. The emission spectrum in figure 3(b) was obtained by exciting the hBN CBG with a 532 nm laser using a high NA objective lens and collecting the emission through the fibre. As shown in the two maps, the spatial distribution of the cavity mode is similar to the simulation result, where the strong cavity-enhanced PL is displayed at the centre of the CBG cavity. Figure 3(c) exemplifies the fibre coupling efficiency. We compared the PL emission collected through the fibre from both the integrated hBN CBG, and a pristine hBN flake. The integrated CBG exhibits a significantly enhanced PL emission (x16) relative to that of a pristine hBN flake.

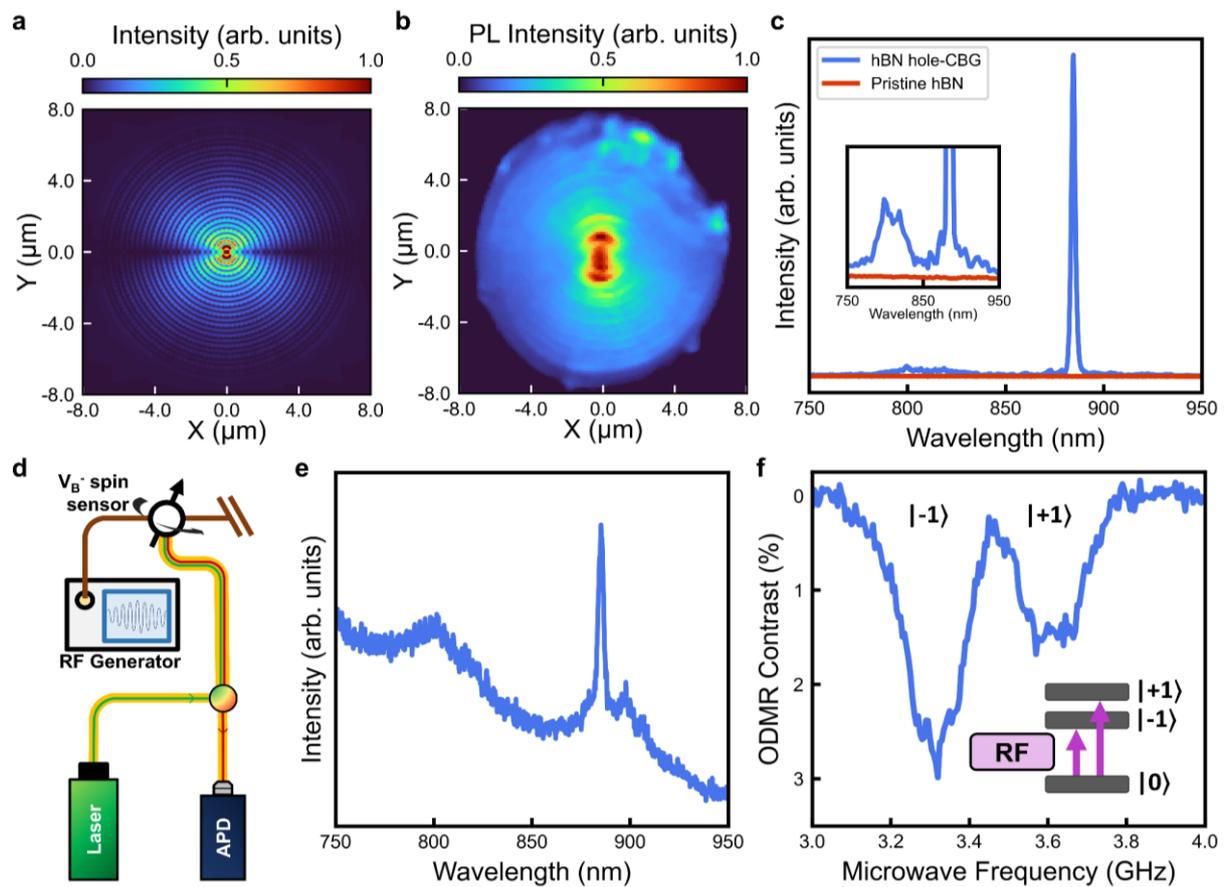

*Figure 3. Optical and spin characterisation of the fibre-integrated hBN CBG devices. (a) FDTD simulation of the near-field power of the hBN CBG cavity, and (b) PL map of the fabricated device on an optical fibre excited using a high NA objective and collected through the fibre. The measured spatial distribution of the cavity mode clearly matches the simulation. (c) The measured PL spectrum from the integrated hBN CBG cavity (blue) and a pristine hBN flake (red). The excitation and collection setup is similar to (b). The CBG mode can be seen alongside the $V_B^-$ emission in the CBG PL spectrum. The inset is a zoomed-in view of the spectra. (d) Simplified schematic of the all-in-fibre excitation and collection configuration. (e) PL spectrum obtained using the all-in-fibre configuration of $V_B^-$ defects in the integrated CBG cavity. (f)*

*Optically detected magnetic resonance (ODMR) measurement of the integrated hBN CBG cavity. Both excitation and collection are done through the fibre. The inset is a simplified energy level diagram of the $V_B^-$ defect ground state (S = 1) split into the three electron spin states: $m_s = 0$, $m_s = \pm 1$.*

Next, we performed both the excitation and collection of the $V_B^-$ emission from the hBN CBG cavity using the fibre, as is shown schematically in Figure 3(d) (i.e. without using an objective lens). Further setup details are provided in Supplementary Information. A PL spectrum from cavity-coupled $V_B^-$ spin defects is shown in Figure 3(e). The distinct sharp peak at 885 nm corresponds to the cavity mode, which dominated the broadband spectrum of the $V_B^-$ emission in the range 750~900 nm. Thus, we can conclude that the cavity-enhanced $V_B^-$ emissions can be efficiently excited and collected through the optical fibre.

A key function of the fibre-integrated hBN device is to detect spin-dependent electron transitions of $V_B^-$ defects by monitoring their PL intensity, known as optically detected magnetic resonance (ODMR). Through the application of a resonant radiofrequency (RF) signal, electrons can be cycled between the $|0\rangle$ and $|\pm 1\rangle$ spin states within the electronic ground state manifold (Inset in Figure 3(f)). The resonant frequency of the spin transitions is highly sensitive to external magnetic fields, and thus enables opportunities for high-resolution, high-sensitivity quantum magnetometry.

We confirmed the sensing functionality by performing ODMR measurements using the fibre-integrated device. The cavity-embedded $V_B^-$ centres were excited and the emission was collected as shown in Figure 3(d). An RF antenna was positioned near the fibre as well as an external magnet to induce a weak magnetic field. The ODMR plot is shown in Figure 3(f) with two distinct transitions at ~ 3.3 GHz and ~ 3.6 GHz, corresponding to the $|-1\rangle$ and $|+1\rangle$ spin states of the $V_B^-$ defect, respectively. The ODMR contrast is approaching 3%, on par with contrast levels recorded from $V_B^-$ ensembles using bulk optics and confocal microscopes[33, 34]. Additional characterisation data is shown in the Supplementary Information.

The central metric for evaluating the performance of a magnetometer is its sensitivity. The magnetic field sensitivity ($\eta_B$) of the fibre-integrated device can be calculated using the following equation:

$$\eta_B \ (T/\sqrt{Hz}) = \mathcal{P}_\mathcal{F} \times \frac{1}{\gamma_e} \times \frac{\Delta\nu_{lw}}{C\sqrt{\mathcal{R}}} \qquad (1)$$

Where the electron gyromagnetic ratio ($\gamma_e$) is equal to 28 MHz/mT and $\mathcal{P}_F$ is a numerical parameter that is determined by the spin resonance shape, 0.70 for a Gaussian profile (See Supplementary Information Section V). The linewidth ($\Delta\nu_{lw}$), ODMR contrast (C) and photon counts per second (R) are all acquired through fitting a Gaussian curve to the ODMR resonant transitions. The calculated magnetic field sensitivity for the fibre-integrated device is 145 µT/√Hz, which is sufficient for many measurements that require remote sensing – e.g., magnetic particle sensing for drug delivery[35] or electrical current detection[35]. We stress, however, that our CBG-cavity integrated fibre magnetometer is fully mobile, without necessity of bulk optics, and offers an excellent signal-to-noise ratio due to its operation in the near infrared spectral range.

To demonstrate potential application of the fibre integrated hBN CBG sensor, we used it to detect remotely layers of steel. Nondestructive measurement of metal thickness and uniformity is of great importance in several industrial fields. As is shown schematically in Figure 4(a), an external magnet was placed above a sheet of steel of increasing thickness. A fibre tip containing the hBN CBG cavity was brought into proximity of the sample and individual ODMR measurements were recorded. The change in magnetic field strength incident on the fibre modifies the magnitude of Zeeman splitting, and induces a Zeeman shift ($\Delta v_Z$), that maps onto a change in magnetic field strength ($B_m$) through the following equation:

$$\Delta v_Z = \frac{g_e \mu_B}{h} \cdot B_m \qquad (2)$$

Requiring no calibration, the transformation is dictated by two fundamental values, the Bohr magneton ($\mu_B$) and Planck's constant ($h$), as well as the dimensionless g-factor ($g_e$), which is approximately equal to 2.0023 for a free electron in a magnetic field. The results of the measurement are shown in Figure 4(b) along with a representation of the shielding effect observed from wide layers of steel. The two representations refer to the cases with no steel (0 mm) and 9 mm of steel. The optical fibre depicted in the inset indicates where the fibre-integrated CBG was located relative to the magnet and the steel. The change in magnetic field strength due to the increasing thickness of steel was measured by the change in resonant frequency relative to 0 mm of steel (i.e. the strength of the external magnet). Using Equation 2, the change in magnetic field strength due to maximum shielding (9 mm) was calculated to be ~10.9 mT.

Finally, we present a unique application of the fibre integrated hBN CBG sensor, which measures magnetic field orientation. This functionality relies on the directional uniformity of the $V_B^-$ magnetic dipoles, produced by the $D_{3h}$ symmetry of the defect, which are aligned out of the plane of hBN[23]. In Figure 4(c), the external magnetic field was rotated from perpendicular to parallel relative to the hBN plane, while the distance was kept constant. The change in magnetic field strength along the $V_B^-$ magnetic dipoles was then measured by the difference in resonant frequency from the perpendicular angle. The strength of the magnetic field vector adheres to a sinusoidal curve and has been fit accordingly. These measurements confirm the successful operation of the fibre-integrated hBN CBG magnetometer without the use of bulk optics.

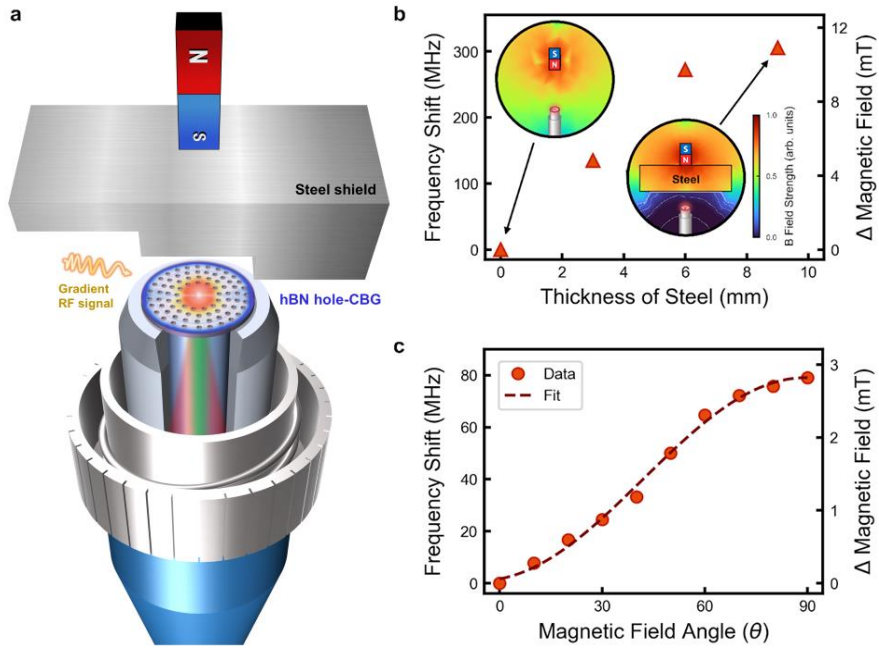

***Figure 4. Remote quantum sensing with the integrated hBN CBG fibre. (a)*** *Configuration for the remote quantum sensing of increasing layers of steel.* ***(b)*** *Plot of change in resonant frequency as a function of steel thickness. The change in magnetic field strength is calculated using Equation 2. The insets show the magnetic field strength incident on the fibre for steel thicknesses of 0 and 9 mm, respectively.* ***(c)*** *An external magnet was rotated around the fibre and a change in resonant frequency was recorded. The values are fit with a sinusoidal curve.*

To conclude, we designed and engineered a new generation of fibre-integrated functional sensing devices. In particular, we leveraged the properties of hBN as a van der Waals crystal that hosts optically-active spin defects, to fabricate monolithic CBG cavities, and transferred them onto the core of an optical fibre. We quantified the performance of the sensor by detecting external layers of metal and magnetic field orientation. This remote-sensing technology can be a good alternative to traditional contact-based ultrasound thickness gauges. Possible orientation-sensitive measurements may include imaging in confined spaces and remote locations where bulk optics aren't suitable. Notably, fibre-integrated optically detectable spin states at room temperature offer distinct advantages for imaging liquids, sensing temperature in extreme environments, nondestructively probing molecules, and detecting micro-faults.

Compared to previously demonstrated fibre-integrated sensors, our approach incorporates a nanophotonic cavity with an optimal optical interface. It is advantageous for fibre excitation and collection, as well as enhances the spatial resolution due to its focusing abilities into sub-micron scanning probes, therefore paving ways to achieve super resolution imaging. The demonstrated design and engineering of hBN CBG cavities can also be extended to other applications, facilitating on-demand, room temperature quantum devices.

## Methods

***Fabrication and Design:*** Finite-difference time-domain (FDTD) simulations were used to optimise nanohole CBG design. The CBG structure is defined by the following four parameters: the radius of the central disk ($R$), the radial interval between nanoholes arrays ($r$), the tangential interval between nanoholes ($l$), and the radius of the nanohole ($a$). Upon optimisation, the ideal parameters of the cavity design are R ~ 645 nm, r ~ 498 nm, l ~ 224 nm, and a ~ 75 nm, designed for an hBN flake with a thickness of 200 nm.

The cavities were patterned using electron beam lithography and transferred into the hBN with fluorine-based reactive ion etching. The hBN flakes were then irradiated with a nitrogen ion beam to create boron vacancies. Devices were then transferred to an etched fibre core with a pattern transfer method using a PDMS stamp. A more in depth fabrication steps can be found in Supplementary Information.

***Optical Measurements:*** Objective excitation and fibre collection measurements as seen in Figure 3(b,c) were performed using a home-built confocal microscope. A 532 nm laser is focused onto the fibre-integrated CBG device by an 100x (0.9 NA) objective. To obtain the confocal PL image, the fibre-integrated hBN device was scanned using an XYZ piezo stage, and the PL emission was collected through the fibre, and coupled into either a spectrometer or avalanche photodiode (APD). All-in-fibre excitation and collection measurements were completed using the home-built optical setup using a 532 nm laser, coupled into the optical fibre. The emission was collected back through the same fibre. Filtering the laser from the emission was done in a free space, using dichroic mirrors and filters

*Optically detected magnetic resonance:* ODMR was performed on the fibre-integrated CBG device placing an RF antenna close to the tip of the fibre. A signal generator swept through a range of RF frequencies, and the photons collected by the APD during each signal and reference pulse were then used to calculate the ODMR contrast.

## Acknowledgements

The authors acknowledge financial support from the Australian Research Council (CE200100010, FT220100053), the Office of Naval Research Global (N62909-22-1-2028), the National Research Foundation of Korea grant funded by MSIT (2022M3H4A1A04096396, 2022R1A2C2003176) and the ITRC program (IITP-2023-2020-0-01606, RS-2023-00259676) for financial support. We thank the UTS node of the ANFF for the access to nanofabrication facilities.

## References

1. Elsherif, M.; Salih, A. E.; Muñoz, M. G.; Alam, F.; AlQattan, B.; Antonysamy, D. S.; Zaki, M. F.; Yetisen, A. K.; Park, S.; Wilkinson, T. D.; Butt, H., Optical Fiber Sensors: Working Principle, Applications, and Limitations. *Advanced Photonics Research* **2022,** *3* (11), 2100371.
2. In *Nonlinear Fiber Optics (Fifth Edition)*, Agrawal, G., Ed. Academic Press: Boston, 2012; pp 621-629.
3. Jauregui, C.; Limpert, J.; Tünnermann, A., High-power fibre lasers. *Nat. Photonics* **2013,** *7* (11), 861-867.

4.	Ngo, G. Q.; Najafidehaghani, E.; Gan, Z.; Khazaee, S.; Siems, M. P.; George, A.; Schartner, E. P.; Nolte, S.; Ebendorff-Heidepriem, H.; Pertsch, T.; Tuniz, A.; Schmidt, M. A.; Peschel, U.; Turchanin, A.; Eilenberger, F., In-fibre second-harmonic generation with embedded two-dimensional materials. *Nat. Photonics* **2022,** *16* (11), 769-776.
5.	Lu, P.; Lalam, N.; Badar, M.; Liu, B.; Chorpening, B. T.; Buric, M. P.; Ohodnicki, P. R., Distributed optical fiber sensing: Review and perspective. *Applied Physics Reviews* **2019,** *6* (4), 041302.
6.	Chen, J.-h.; Xiong, Y.-f.; Xu, F.; Lu, Y.-q., Silica optical fiber integrated with two-dimensional materials: towards opto-electro-mechanical technology. *Light: Science & Applications* **2021,** *10* (1), 78.
7.	Wang, Z.; Wang, G.; Kumar, S.; Marques, C.; Min, R.; Li, X., Recent Advancements in Resonant Fiber Optic Gyro—A Review. *IEEE Sensors Journal* **2022,** *22* (19), 18240-18252.
8.	Luo, Z.; Li, M.; Kong, X.; Li, Y.; Li, W.; Tian, Z.; Cao, Q.; Zaman, M. H.; Li, Y.; Xiao, W.; Duan, Y., Advance on fiber optic-based biosensors for precision medicine: From diagnosis to therapy. *Interdisciplinary Medicine* **2023,** *1* (4), e20230022.
9.	Jani, M.; Czarnecka, P.; Orzechowska, Z.; Mrózek, M.; Gawlik, W.; Wojciechowski, A. M., Sensing of Magnetic-Field Gradients with Nanodiamonds on Optical Glass-Fiber Facets. *ACS Applied Nano Materials* **2023,** *6* (13), 11077-11084.
10.	Chen, Y.; Lin, Q.; Cheng, H.; Huang, H.; Shao, J.; Ye, Y.; Liu, G.-S.; Chen, L.; Luo, Y.; Chen, Z., Nanodiamond-Based Optical-Fiber Quantum Probe for Magnetic Field and Biological Sensing. *ACS Sensors* **2022,** *7* (12), 3660-3670.
11.	Scholten, S. C.; Healey, A. J.; Robertson, I. O.; Abrahams, G. J.; Broadway, D. A.; Tetienne, J. P., Widefield quantum microscopy with nitrogen-vacancy centers in diamond: Strengths, limitations, and prospects. *J. Appl. Phys.* **2021,** *130* (15), 150902.
12.	Rondin, L.; Tetienne, J. P.; Hingant, T.; Roch, J. F.; Maletinsky, P.; Jacques, V., Magnetometry with nitrogen-vacancy defects in diamond. *Reports on Progress in Physics* **2014,** *77* (5), 056503.
13.	Ebendorff-Heidepriem, H.; Ruan, Y.; Ji, H.; Greentree, A. D.; Gibson, B. C.; Monro, T. M., Nanodiamond in tellurite glass Part I: origin of loss in nanodiamond-doped glass. *Optical Materials Express* **2014,** *4* (12), 2608-2620.
14.	Stürner, F. M.; Brenneis, A.; Buck, T.; Kassel, J.; Rölver, R.; Fuchs, T.; Savitsky, A.; Suter, D.; Grimmel, J.; Hengesbach, S.; Förtsch, M.; Nakamura, K.; Sumiya, H.; Onoda, S.; Isoya, J.; Jelezko, F., Integrated and Portable Magnetometer Based on Nitrogen-Vacancy Ensembles in Diamond. *Advanced Quantum Technologies* **2021,** *4* (4), 2000111.
15.	Zhao, M.; Lin, Q.; Meng, Q.; Shan, W.; Zhu, L.; Chen, Y.; Liu, T.; Zhao, L.; Jiang, Z., All Fiber Vector Magnetometer Based on Nitrogen-Vacancy Center. *Nanomaterials (Basel, Switzerland)* **2023,** *13* (5).
16.	Li, Y.; Gerritsma, F. A.; Kurdi, S.; Codreanu, N.; Gröblacher, S.; Hanson, R.; Norte, R.; van der Sar, T., A Fiber-Coupled Scanning Magnetometer with Nitrogen-Vacancy Spins in a Diamond Nanobeam. *ACS Photonics* **2023,** *10* (6), 1859-1865.
17.	Patel, R. N.; Schröder, T.; Wan, N.; Li, L.; Mouradian, S. L.; Chen, E. H.; Englund, D. R., Efficient photon coupling from a diamond nitrogen vacancy center by integration with silica fiber. *Light: Science & Applications* **2016,** *5* (2), e16032-e16032.
18.	Liu, H.-Y.; Liu, W.-Z.; Wang, M.-Q.; Ye, X.-Y.; Yu, P.; Sun, H.-Y.; Liu, Z.-X.; Liu, Z.-X.; Zhou, J.-W.; Wang, P.-F.; Shi, F.-Z.; Wang, Y., Nanoscale Vector Magnetometry with a Fiber-Coupled Diamond Probe. *Advanced Quantum Technologies* **2023,** *6* (9), 2300127.

19. Vaidya, S.; Gao, X.; Dikshit, S.; Aharonovich, I.; Li, T., Quantum sensing and imaging with spin defects in hexagonal boron nitride. *Advances in Physics: X* **2023,** *8* (1), 2206049.
20. Caldwell, J. D.; Aharonovich, I.; Cassabois, G.; Edgar, J. H.; Gil, B.; Basov, D. N., Photonics with hexagonal boron nitride. *Nature Reviews Materials* **2019,** *4* (8), 552-567.
21. Aharonovich, I.; Tetienne, J.-P.; Toth, M., Quantum Emitters in Hexagonal Boron Nitride. *Nano Lett.* **2022,** *22* (23), 9227-9235.
22. Healey, A. J.; Scholten, S. C.; Yang, T.; Scott, J. A.; Abrahams, G. J.; Robertson, I. O.; Hou, X. F.; Guo, Y. F.; Rahman, S.; Lu, Y.; Kianinia, M.; Aharonovich, I.; Tetienne, J. P., Quantum microscopy with van der Waals heterostructures. *Nature Phys.* **2022**.
23. Gottscholl, A.; Kianinia, M.; Soltamov, V.; Orlinskii, S.; Mamin, G.; Bradac, C.; Kasper, C.; Krambrock, K.; Sperlich, A.; Toth, M.; Aharonovich, I.; Dyakonov, V., Initialization and read-out of intrinsic spin defects in a van der Waals crystal at room temperature. *Nature Mater.* **2020,** *19* (5), 540-545.
24. Gottscholl, A.; Diez, M.; Soltamov, V.; Kasper, C.; Krauße, D.; Sperlich, A.; Kianinia, M.; Bradac, C.; Aharonovich, I.; Dyakonov, V., Spin defects in hBN as promising temperature, pressure and magnetic field quantum sensors. *Nat. Commun.* **2021,** *12* (1), 4480.
25. Gao, X.; Vaidya, S.; Li, K.; Ju, P.; Jiang, B.; Xu, Z.; Allcca, A. E. L.; Shen, K.; Taniguchi, T.; Watanabe, K.; Bhave, S. A.; Chen, Y. P.; Ping, Y.; Li, T., Nuclear spin polarization and control in hexagonal boron nitride. *Nature Mater.* **2022**, DOI: 10.1038/s41563-022-01329-8.
26. Huang, M.; Zhou, J.; Chen, D.; Lu, H.; McLaughlin, N. J.; Li, S.; Alghamdi, M.; Djugba, D.; Shi, J.; Wang, H.; Du, C. R., Wide field imaging of van der Waals ferromagnet $Fe_3GeTe_2$ by spin defects in hexagonal boron nitride. *Nat. Commun.* **2022,** *13* (1), 5369.
27. Sasaki, K.; Nakamura, Y.; Gu, H.; Tsukamoto, M.; Nakaharai, S.; Iwasaki, T.; Watanabe, K.; Taniguchi, T.; Ogawa, S.; Morita, Y.; Kobayashi, K., Magnetic field imaging by hBN quantum sensor nanoarray. *Appl. Phys. Lett.* **2023,** *122* (24), 244003.
28. Fröch, J. E.; Spencer, L. P.; Kianinia, M.; Totonjian, D. D.; Nguyen, M.; Gottscholl, A.; Dyakonov, V.; Toth, M.; Kim, S.; Aharonovich, I., Coupling Spin Defects in Hexagonal Boron Nitride to Monolithic Bullseye Cavities. *Nano Lett.* **2021,** *21* (15), 6549-6555.
29. Qian, C.; Villafañe, V.; Schalk, M.; Astakhov, G. V.; Kentsch, U.; Helm, M.; Soubelet, P.; Wilson, N. P.; Rizzato, R.; Mohr, S.; Holleitner, A. W.; Bucher, D. B.; Stier, A. V.; Finley, J. J., Unveiling the Zero-Phonon Line of the Boron Vacancy Center by Cavity-Enhanced Emission. *Nano Lett.* **2022,** *22* (13), 5137-5142.
30. Li, P.; Dolado, I.; Alfaro-Mozaz, F. J.; Casanova, F.; Hueso, L. E.; Liu, S.; Edgar, J. H.; Nikitin, A. Y.; Vélez, S.; Hillenbrand, R., Infrared hyperbolic metasurface based on nanostructured van der Waals materials. *Science* **2018,** *359* (6378), 892-896.
31. Castellanos-Gomez, A.; Buscema, M.; Molenaar, R.; Singh, V.; Janssen, L.; van der Zant, H. S. J.; Steele, G. A., Deterministic transfer of two-dimensional materials by all-dry viscoelastic stamping. *2D Materials* **2014,** *1* (1), 011002.
32. Jeon, W. B.; Moon, J. S.; Kim, K.-Y.; Ko, Y.-H.; Richardson, C. J. K.; Waks, E.; Kim, J.-H., Plug-and-Play Single-Photon Devices with Efficient Fiber-Quantum Dot Interface. *Advanced Quantum Technologies* **2022,** *5* (10), 2200022.
33. Liang, H.; Chen, Y.; Yang, C.; Watanabe, K.; Taniguchi, T.; Eda, G.; Bettiol, A. A., High Sensitivity Spin Defects in hBN Created by High-Energy He Beam Irradiation. *Adv. Opt. Mater.* **2023,** *11* (1), 2201941.

34. Xu, X.; Solanki, A. B.; Sychev, D.; Gao, X.; Peana, S.; Baburin, A. S.; Pagadala, K.; Martin, Z. O.; Chowdhury, S. N.; Chen, Y. P.; Taniguchi, T.; Watanabe, K.; Rodionov, I. A.; Kildishev, A. V.; Li, T.; Upadhyaya, P.; Boltasseva, A.; Shalaev, V. M., Greatly Enhanced Emission from Spin Defects in Hexagonal Boron Nitride Enabled by a Low-Loss Plasmonic Nanocavity. *Nano Lett.* **2023,** *23* (1), 25-33.
35. Herea, D.-D.; Lăbuşcă, L.; Lupu, N.; Chiriac, H., 10 - Magnetic particles for drug delivery. In *Magnetic Sensors and Actuators in Medicine*, Chiriac, H.; Lupu, N., Eds. Woodhead Publishing: 2023; pp 259-304.